\documentclass[9pt,twocolumn,twoside]{osajnl}

\journal{ao} 

\setboolean{shortarticle}{false}
\usepackage{soul}

\title{Balanced photodetectors nonlinearity for short-pulse regime}

\author[1,*]{Philippe Guay}
\author[1]{Jérôme Genest}

\affil[1]{Centre d'optique, photonique et laser, Universit\'{e} Laval, Qu\'{e}bec, Qu\'{e}bec G1V 0A6, Canada}

\affil[*]{Corresponding author: philippe.guay.4@ulaval.ca}

\begin{abstract}
Short pulse lasers are used to characterize the nonlinear response of amplified photodetectors. Two widely used balanced detectors are characterized in terms of amplitude, area, broadening, and balancing the mismatch of their impulse response. The dynamic impact of pulses on the detector is also discussed. It is demonstrated that using photodetectors with short pulses triggers nonlinearities even when the source average power is well below the detector continuous power saturation threshold. 
\end{abstract}

\setboolean{displaycopyright}{true}

\begin{document}

\maketitle



Commercially available balanced photodetectors are widely used as they allow significant improve in signal-to-noise ratio (SNR) of a measurement by suppressing common-mode fluctuations between their inputs. Balanced detectors have been routinely used with short pulses in applications such as optical time-domain interferometry \cite{LU12,SOR21,TU15,ZHA16}, dual-comb spectroscopy \cite{GUA19,FDI19,HEB17,ZHAO16,CHE19}, metrological instrumentation \cite{KIM14,BEN12}, nondestructive testing \cite{PEL14}, LIDAR \cite{DIA14}, nonlinear optics \cite{CHE15,SHI18}, Kerr rotation microscopy \cite{HUA17}, optical coherent tomography \cite{EOM08}, and so on. 

Although short pulses are widely used in optics, their detection has always been challenging. The high energy content of the pulses can momentarily saturate a photodetector (PD) even though the equivalent continuous wave (CW) power level is well below the CW saturation level of the detector. Even with low energy pulses and average power level far below the CW saturation level, the peak power levels reached can greatly exceed the saturation level. Very often, measurements are negatively impacted by the nonlinear response of detectors. Reducing power onto the detector is an easy solution to the problem. However, it often is advantageous to operate a photodetector in a high power mode because SNR increases linearly with power, while it increases only with the square root of measurement time \cite{ROY12}. As a result, balanced detectors are often used with short pulses sources having average powers near the CW saturation limit. Thus, one has to be mindful of the impact of high intensity pulses on photodetectors. Understanding the nonlinear behavior of balanced detectors can help improve their design and may enable better nonlinearity (NL) handling in short-pulse precision measurements. 



In this letter, we present an extended characterization of two widely used balanced photodetectors to demonstrate the imperfect balancing between photodiodes caused by the uneven broadening of the PD's impulse responses, the mismatched ringing of the impulse responses, and the dynamic impact of pulse saturation. Thorlabs' PDB130C and PDB480C amplified detectors having respectively a 350 MHz and a 1.6 GHz bandwidth are used for this demonstration. An overview of the impulse response of a PDB430C is also provided to support the generalization of PD nonlinearity in Thorlabs' PDB series. Measurements presented here point to the amplifing stages, rather than photodiodes themselves being the main NL drivers. This provides hints that efforts towards a more adapted electronic design as was done in \cite{HOB97} for noise cancellation circuits could provide greatly enhanced performance.  


\begin{figure}[htbp]
\centering
\includegraphics[width=\linewidth]{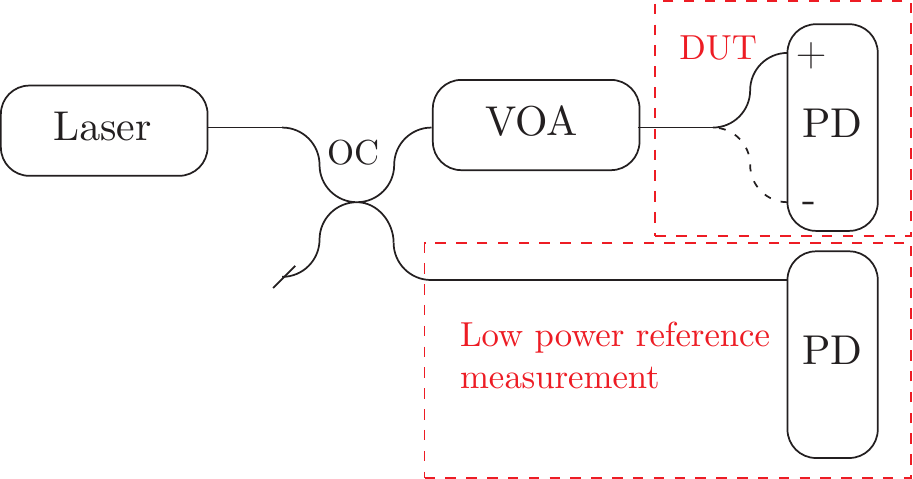}
\caption{Simplified block diagram of the photodetectors characterization. OC: Optical coupler. VOA: Variable optical attenuator. DUT: Device under test. PD: Photodetector}
\label{fig:setup}
\end{figure}

The experimental setup for the photodetector characterization is shown in Fig. \ref{fig:setup}. In the case of the PDB130C, the laser used was a homemade femtosecond mode-locked source based on nonlinear polarization rotation (NPR) \cite{GIA08}. The source has a repetition rate of 20 MHz to make sure individual impulse responses are well separated. The laser was split to study the device under test (DUT) while simultaneously providing a low power reference. The reference arm was used to properly align the high power impulse responses with respect to a low power input that is in the same condition for all measurements. In the DUT arm, a variable optical attenuator was used to vary the power sent to the photodiode. The power measured by the attenuor was switched between the positive and negative photodiodes of the detector to separately measure the impulse response of both branches. 

\begin{figure}[htbp]
\centering
\includegraphics[width=\linewidth]{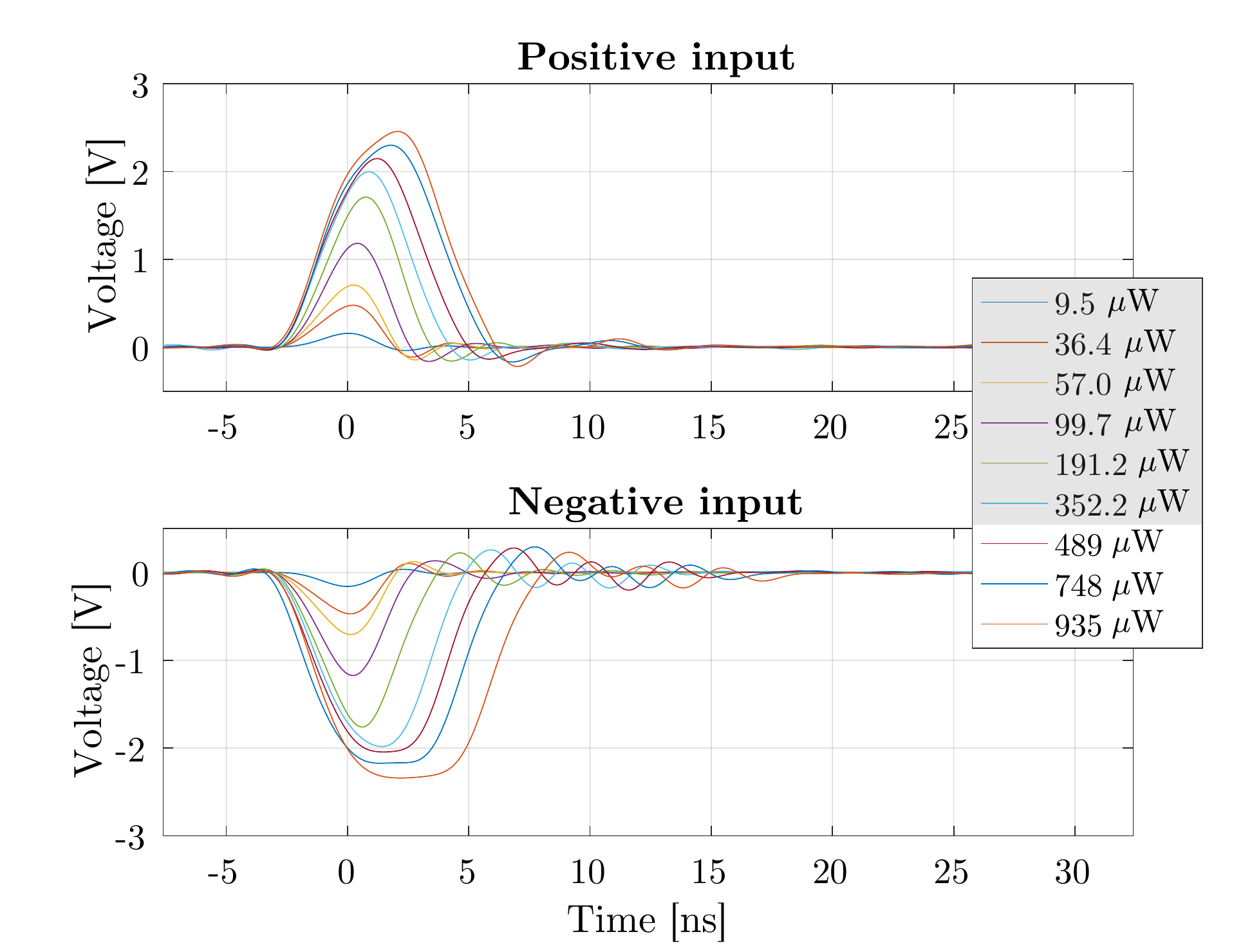}
\caption{Impulse responses for the positive and the negative photodiodes of Thorlabs' balanced detector PDB130C. Power levels below the CW saturation threshold are highlighted in the legend.}
\label{fig:pulses_PDB130C}
\end{figure}

The positive and negative impulse responses of the PDB130C are shown in Fig. \ref{fig:pulses_PDB130C} for various incident power levels where the levels under the CW saturation of the PD (<400 $\mu$W) have been highlighted. Given that the repetition rate of the laser is less than a tenth of the detector's bandwidth, the impulse responses are well separated and do not overlap each other. The responses have been aligned using a time grid provided by the reference signal. Since the photodetector is AC-coupled, an offset corresponding the DC level observed between the pulses has been added to the traces shown in Fig. \ref{fig:pulses_PDB130C} for comparison purposes. Because the power level is constant on the detector and successive pulses are similar from the photodetector standpoint, this effectively removes the high-pass filter impact.

\begin{figure}[htbp]
\centering
\includegraphics[width=\linewidth]{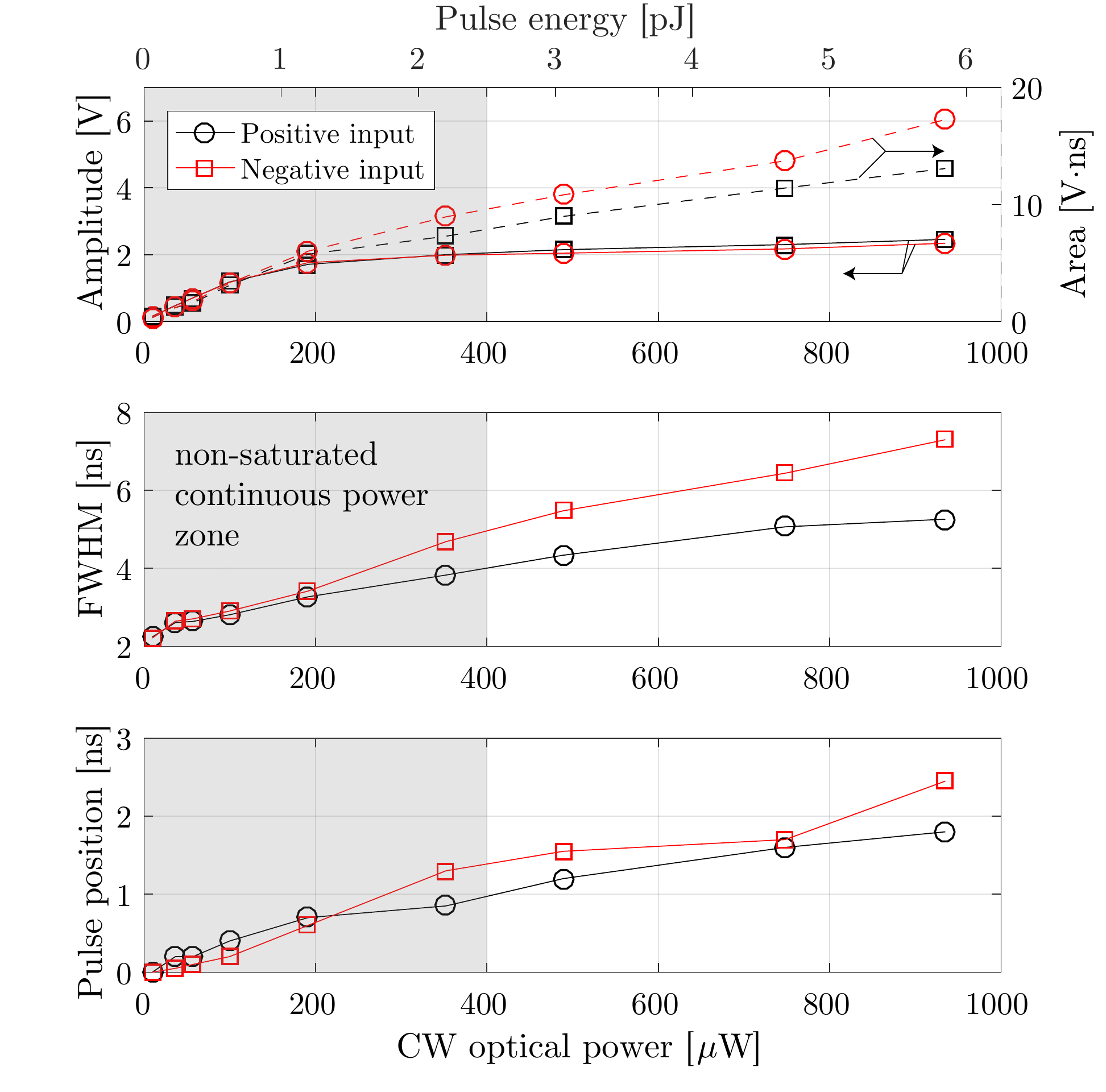}
\caption{Extracted parameters for the impulses responses. The amplitude, area, full-width-half-maximum and the pulse position are shown as a function of the incident power level. \label{fig:characterizationPDB130C}}
\end{figure}

The response amplitude and area as a function of the incident power are shown on the top panel of Fig. \ref{fig:characterizationPDB130C}. The saturation of the amplitude is clear as the curve reaches a plateau around 2 V for both photodiodes, while the area continues to increase somewhat linearly with a different slope for both photodiodes. Since the amplitude of the impulse responses is saturating, but the area continues to increase with power, the width of the pulses increases as a corollary. This effect is seen in the middle panel of Fig. \ref{fig:characterizationPDB130C} where the pulse full width half maximum (FWHM) is displayed as a function of power. The figure shows that the rate at which the FWHM increases is different for both photodiodes. As a result of this time broadening, the position of the pulses moves with increasing power. This can be viewed as a dynamic impact of the nonlinear response of the detector. The bottom panel of Fig. \ref{fig:characterizationPDB130C} shows the position of the pulses calculated from the average position between the rising edge and the falling edge at 90\% of the maximum.

For all the parameters reported in Fig. \ref{fig:characterizationPDB130C}, including the ringing of the impulse responses, diverging behaviors are observed between the photodiodes. The disagreement is even observed for  power levels significantly lower than the CW saturation level provided by the manufacturer. This implies that matching conditions for the positive and negative channels of the photodetectors could be improved.

The observations made with a PDB130C have been validated using a PDB430C having similar characteristics such as bandwidth and gain.  The general behaviour is broadly similar (amplitude saturation, pulse broadening and ringing), but specific values vary greatly from one photodetector to another.



\begin{figure}[htbp]
\centering
\includegraphics[width=\linewidth]{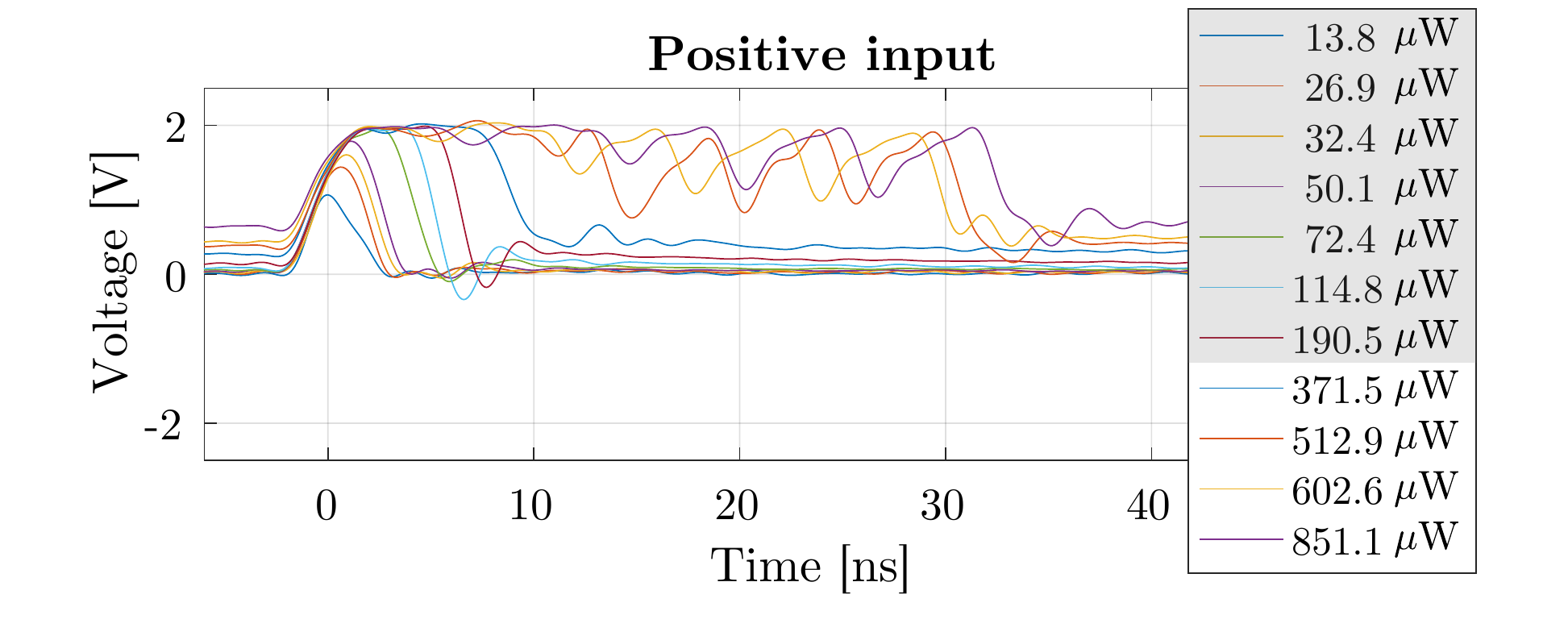}
\caption{Impulse responses for the positive photodiode of Thorlabs' balanced detector PDB430C. Power levels below the CW saturation threshold are highlighted in the legend.}
\label{fig:pulses_PDB430C}
\end{figure}

Fig. \ref{fig:pulses_PDB430C} shows the impulse response of a PDB430C (positive input only) illustrating the general trend observed so far, but with a particularity : the output reaches an apparently chaotic state beyond 370 $\mu$W. 

Thorlabs' PDB480C has also been studied. This balanced detector has a 1.6 GHz bandwidth which enables use of a source having a faster repetition rate, while maintaining pulse independence. Here, a 100 MHz repetition rate Menlo c-comb was used. Again, the low repetition rate of the laser compared to the photodetector's bandwidth allows resolving individual impulse responses, ensuing that the latter are independent. 

\begin{figure}[htbp]
\centering
\includegraphics[width=\linewidth]{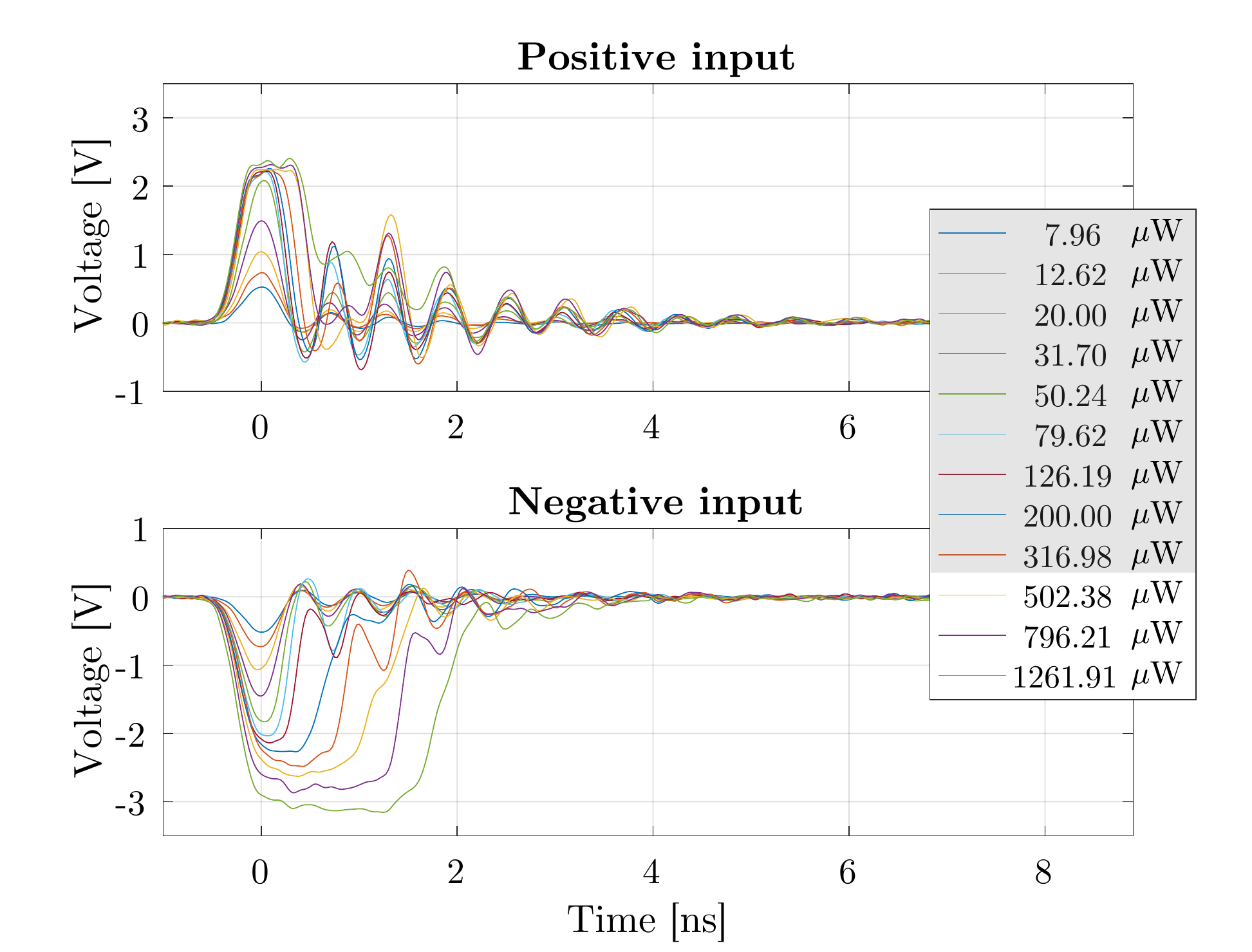}
\caption{Impulse responses for the positive and the negative photodiodes of Thorlabs' balanced detector PDB40C. Power levels below the CW saturation threshold are highlighted in the legend.}
\label{fig:pulses_PDB480C}
\end{figure}

Fig. \ref{fig:pulses_PDB480C} illustrates the impulse responses of the photodetector's positive and negative inputs. The same analysis used with the PDB130C has been conducted and the conclusions are somewhat similar. Amplitude saturation, increasing area, increasing FWHM and shifted pulses are the most notable features observed. In addition, ringing is quite prominent for the positive photodiode. 

\begin{figure}[htbp]
\centering
\includegraphics[width=\linewidth]{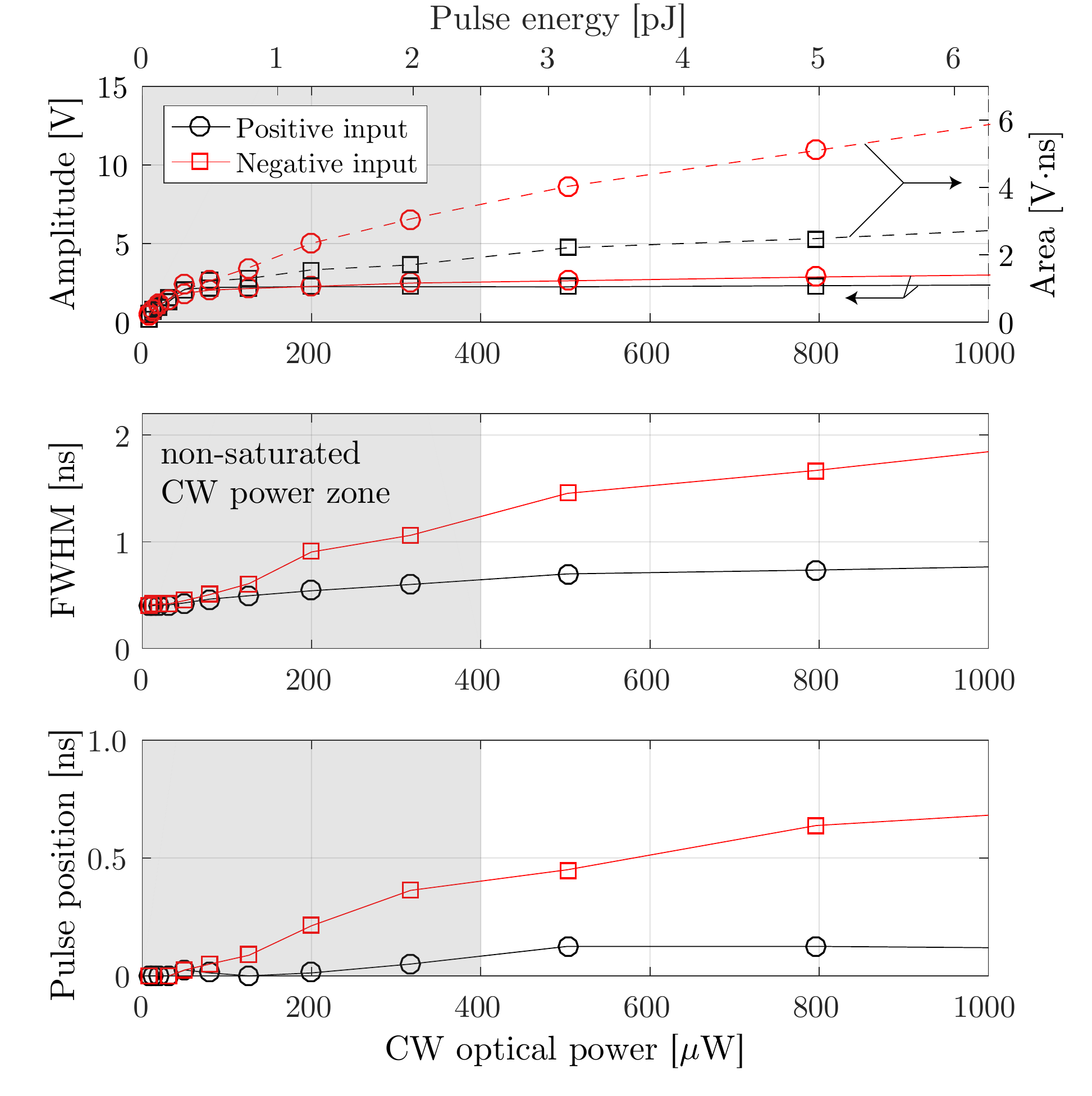}
\caption{(top panel) Full width half maximum (FWHM) of the impulse responses and position of the maximum value of the impulse response (bottom panel) as a function of optical power for the PDB130C detector.}
\label{fig:characterizationPDB480C}
\end{figure}

Furthermore, Fig. \ref{fig:characterizationPDB480C} highlights the diverging behavior between the photodiodes which is more significant than for the PDB130C. For instance, the FWHM for the negative input becomes twice what is seen for the positive input, even below the CW saturation level. 

The analysis of the PDB detectors' impulse responses has shown that the shape of the impulse responses changes as optical power is increased. As saturation is reached, the impulse response stretches towards the next pulse. This becomes problematic when different impulse responses within a varying amplitude signal start to leak on one another. Thus, a deformation of the signal is introduced. For example, a typical 200 MHz frequency comb generated from a pulsed source \cite{SIN14} has a period of 5 ns. In theory, a PDB130C has sufficient bandwidth to properly detect a modulation on such a comb. In practice, however, pulses corresponding to higher intensities may broaden sufficiently to overlap with successive pulses, while low intensity pulses would completely decay before the next one. This means that each pulse detected could have a different weighted contribution from the previous one, thus adding an additional layer of complexity to managing nonlinearities.

\begin{figure}[htbp]
\centering
\includegraphics[width=\linewidth]{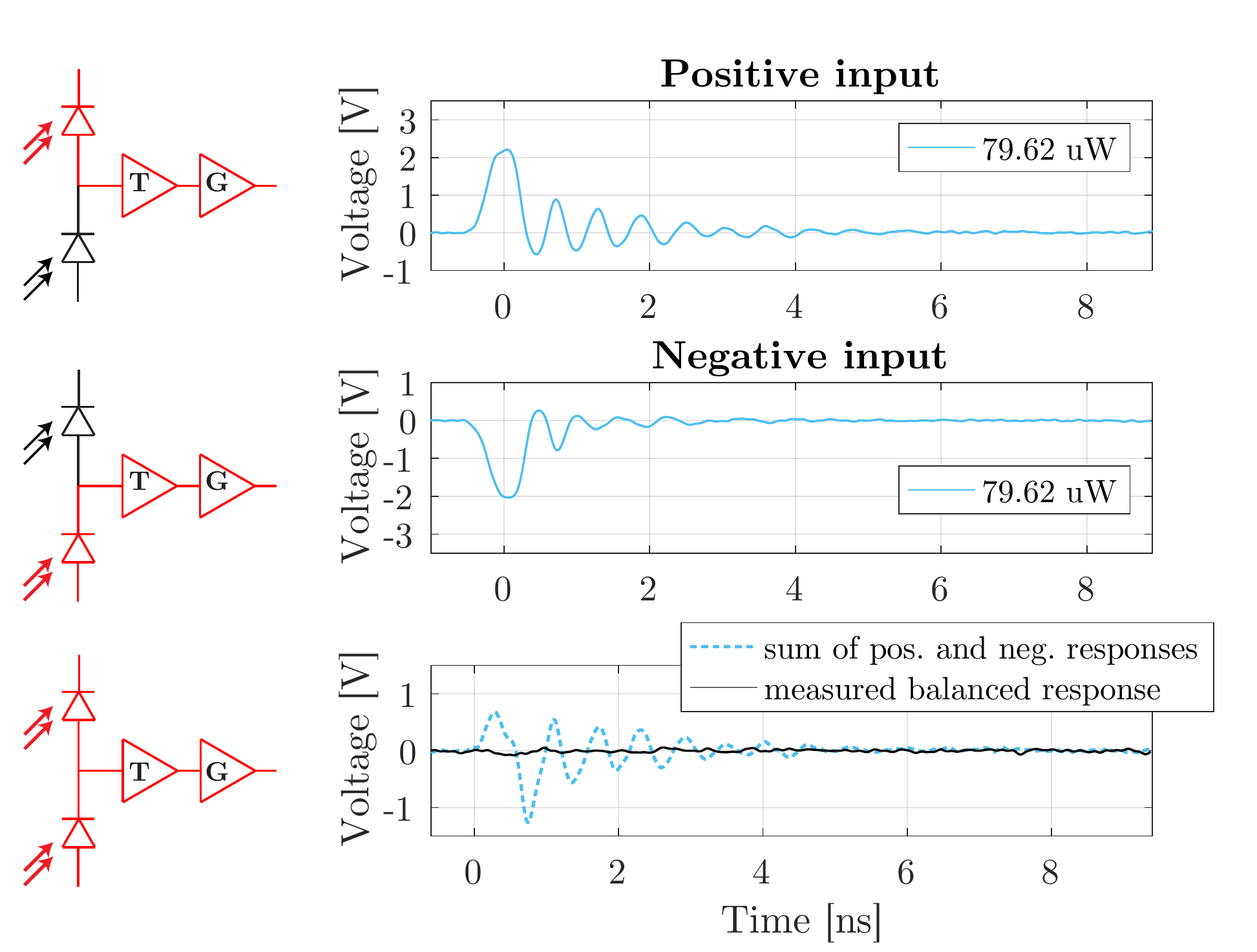}
\caption{Impulse response for the illumination of the positive input (top panel), the negative input (middle panel) and for both input simultaneously (bottom panel). Each plot is sided with the circuit where the flow of current is highlighted in red. T: Transimpedance amplifier. G: Gain amplifier.}
\label{fig:balancedPDB480C}
\end{figure}

One may question the source of this disparity between the positive and negative impulse responses. The photodiodes seem to be the obvious culprits, as they appear to be the main source of asymmetry in the simplified electrical circuit (shown on the left of Fig. \ref{fig:balancedPDB480C}) by providing the added positive and negative currents. However, nonlinearity can also occur in the circuitry following the photodiode. In fact, the current produced by the photodiodes is sent to amplifiers that are likely to exhibit nonlinear behavior notably rail to rail saturation as is observed here. 

To gain insight as to the origin of NL in the circuit, the response to a balanced excitation is compared to the sum of the responses when each photodiode is illuminated separately. If the measured response matches the sum of the responses, one would conclude that the nonlinearity occurs prior to the current sum, hence in the photodiodes. Both currents produced by the photodiodes would have nonlinear characteristics and the sum would then be amplified just as when each current is amplified separately. However, if the measured response does not match the sum of the responses, the nonlinearity occurs after the current sum, most likely in the amplifiers. In this case, both currents produced by the photodiodes would largely cancel each other and there would only be a small signal to amplify. 

As can be seen in Fig. \ref{fig:balancedPDB480C}, a power of 79.62 $\mu$W was sent to the positive photodiode and the impulse response was measured. The same measurement was done for the negative input. The sum of these responses is plotted in a dashed blue line in the bottom panel of Fig. \ref{fig:balancedPDB480C}. This expected response does not match the actual response that was measured when the same optical power was sent to both photodiodes simultaneously. This confirms that the nonlinearity actually occurs after the currents produced by the photodiodes are summed, hence in the amplifiers, most likely in the gain amplifier as Thorlabs' documentation \cite{THO07} indicates that the amplifiers are built from a transimpedance followed by a gain amplifier. 

It turns out that the OPA847 amplifier used in these detectors has an asymmetric recovery from positive or negative saturation (see bottom figures of page 7 in \cite{OPA84}]). This behaviour is compatible with what is observed here. 
 
As a conclusion, an analysis of the impulse responses of two balanced detectors from Thorlabs' PDB line has been provided as well a brief overview of a third detector. It has been shown that one has to be mindful of the nonlinear behavior of the detector when used in pulsed regime. Nonlinear responses include a saturation of the amplitude, a broadening of the pulses as well as a dynamic shift of the responses, even well below the continuous power saturation level quoted by the manufacturer. Observations reported here open the way to the design of better detection electronics. The work may also allow improving the quality of measurements made with unmodified commercial balanced photodetectors by giving means to take properly into account the dynamic nonlinear behavior since the characterization of photodetectors will enable post-processing correction of nonlinearity.      

\begin{backmatter}
\bmsection{Funding} King Abdullah University of Science and Technology (OSR-CRG2019-4046); Fonds de recherche du Québec – Nature et technologies; Natural Sciences and Engineering Research Council of Canada.



\bmsection{Disclosures} The authors declare no conflicts of interest.

\bmsection{Data Availability Statement} Data underlying the results presented in this paper are not publicly available at this time but may be obtained from the authors upon reasonable request.










\end{backmatter}

\bigskip

\bibliography{sample}

\bibliographyfullrefs{sample}


\ifthenelse{\equal{\journalref}{aop}}{%
\section*{Author Biographies}
\begingroup
\setlength\intextsep{0pt}
\begin{minipage}[t][6.3cm][t]{1.0\textwidth} 
  \begin{wrapfigure}{L}{0.25\textwidth}
    \includegraphics[width=0.25\textwidth]{john_smith.eps}
  \end{wrapfigure}
  \noindent
  {\bfseries John Smith} received his BSc (Mathematics) in 2000 from The University of Maryland. His research interests include lasers and optics.
\end{minipage}
\begin{minipage}{1.0\textwidth}
  \begin{wrapfigure}{L}{0.25\textwidth}
    \includegraphics[width=0.25\textwidth]{alice_smith.eps}
  \end{wrapfigure}
  \noindent
  {\bfseries Alice Smith} also received her BSc (Mathematics) in 2000 from The University of Maryland. Her research interests also include lasers and optics.
\end{minipage}
\endgroup
}{}

\end{document}